\documentclass[12pt,preprint]{aastex}
\newcommand{\ergseccm}{erg\,s$^{-1}$\,cm$^{-2}$}
\newcommand{\degree}{\hbox{$^\circ$}}
\newcommand{\etal}{et\,al.}
\newcommand{\halpha}{H$\alpha$}
\newcommand{\lsim}{\raise0.3ex\hbox{$<$}\kern-0.75em{\lower0.65ex\hbox{$\sim$}}}
\newcommand{\mjypbm}{\,\,mJy\,beam$^{-1}$}
\newcommand{\msun}{M$_{\odot}$}
\newcommand{\HII}{H~{\sc ii}}
\newcommand{\HI}{H~{\sc i}}
\newcommand{\kms}{km\,s$^{-1}$}
\newcommand{\ks}{K$_{\rm S}$}
\newcommand{\pom}{$\pm$}
\newcommand{\gsim}{\raise0.3ex\hbox{$>$}\kern-0.75em{\lower0.65ex\hbox{$\sim$}}}
\begin{document}

\slugcomment{Astronomical Journal, in press}

\title{The Stellar and Gaseous Contents of the Orion Dwarf Galaxy}




\author{John M. Cannon, Korey Haynes, Hans Most}
\affil{Department of Physics \& Astronomy, Macalester College, 1600 Grand 
Avenue\\ Saint Paul, MN 55105}
\email{jcannon@macalester.edu}

\author{John J. Salzer}
\affil{Astronomy Department, Wesleyan University, Middletown, CT 06457}
\affil{Department of Astronomy, Indiana University, 727 East Third Street,
Bloomington, IN 47405}

\author{Kaitlin Haugland}
\affil{Griffith Observatory, 2800 East Observatory Road, Los Angeles, CA 90027}
\affil{Department of Physics \& Astronomy, Macalester College, 1600 Grand 
Avenue, Saint Paul, MN 55105}

\author{Jillian Scudder}
\affil{Department of Physics \& Astronomy, University of Victoria, PO Box 3055, STN CSC, Victoria, BC, V8W 3P6 Canada}
\affil{Department of Physics \& Astronomy, Macalester College, 1600 Grand 
Avenue, Saint Paul, MN 55105}

\author{Arthur Sugden}
\affil{Department of Biology and Medicine, Brown University, Providence, RI 02912} 
\affil{Department of Astronomy, Indiana University, 727 East Third Street, 
Bloomington, IN 47405}

\author{Jacob Weindling}
\affil{Department of Physics \& Astronomy, Macalester College, 1600 Grand 
Avenue, Saint Paul, MN 55105}

\begin{abstract}

We present new {\it KPNO 0.9-m} optical and {\it VLA} \HI\ spectral
line observations of the Orion dwarf galaxy.  This nearby (D $\simeq$
5.4 Mpc), intermediate-mass (M$_{\rm dyn} \simeq$
1.1\,$\times$\,10$^{10}$ \msun) dwarf displays a wealth of structure
in its neutral ISM, including three prominent ``hole/depression''
features in the inner \HI\ disk.  We explore the rich gas kinematics,
where solid-body rotation dominates and the rotation curve is flat out
to the observed edge of the \HI\ disk ($\sim$6.8 kpc).  The Orion
dwarf contains a substantial fraction of dark matter throughout its
disk: comparing the 4.7\,$\times$\,10$^{8}$ \msun\ of detected neutral
gas with estimates of the stellar mass from optical and near-infrared
imaging (3.7\,$\times$\,10$^{8}$ \msun) implies a mass-to-light ratio
$\simeq$13.  New \halpha\ observations show only modest-strength
current star formation ($\sim$0.04 \msun\,yr$^{-1}$); this star
formation rate is consistent with our 1.4 GHz radio continuum
non-detection.

\end{abstract}						


\section{Introduction}
\label{S1}

Dwarf galaxies are excellent laboratories for studying the processes
that shape galaxy evolution in canonically ``simple'' conditions.
They lack the rotational shear intrinsic to spirals, meaning that star
formation (SF) is initiated primarily by local mechanisms.  Further,
the structures in the ISM are not immediately destroyed, allowing one
to study the interplay between SF and the ISM.  Targeted observations
and larger scale surveys have demonstrated the diagnostic power of
\HI\ spectral line imaging of nearby dwarf systems (e.g., {Puche
\etal\ 1992}\nocite{puche92}; {Puche \& Westpfahl
1994}\nocite{puche94}; {Hunter \etal\ 2007}\nocite{hunter07}; {Oh
\etal\ 2008}\nocite{oh08}; {Walter \etal\ 2008}\nocite{walter08}).

Nearby, gas-rich dwarf galaxies also offer comparatively ``simple''
environments in which to perform detailed kinematic analysis.
Many dwarfs display solid-body rotation that is well-suited to
precision rotation curve work (e.g., {Weldrake \etal\ 2003}).
Using simple dynamical arguments, most nearby systems appear to be
dark-matter dominated \citep{mateo98}, making them important
laboratories for understanding the missing baryons problem.

Numerous investigations of nearby systems have revealed rich kinematic
structure in the form of shells and holes in the neutral gas (e.g.,
{Kim \etal\ 1999}\nocite{kim99}; {Walter \& Brinks
1999}\nocite{walter99}, {2001}\nocite{walter01}; {Ott \etal\
2001}\nocite{ott01}; to name just a few).  It is commonly assumed that
these structures result from ``feedback'' processes (i.e., stellar
winds and SNe; e.g., {Tenorio-Tagle \& Bodenheimer
1988}\nocite{tenoriotagle88}).  However, debate continues on whether
all such features arise from simple stellar evolution processes (e.g.,
{Rhode \etal\ 1999}\nocite{rhode99}; {S{\' a}nchez-Salcedo
2002}\nocite{sanchezsalcedo02}).

Here, we present new \HI\ and optical observations of the Orion dwarf
galaxy.  Originally discovered by \citet{giovanelli79}, subsequent
{\it Arecibo} observations showed it to be \HI-rich
\citep{springob05}. However, the low Galactic latitude
($-$12.3\arcdeg) and high associated foreground reddening (A$_{\rm R}
=$ 1.959 mag; {Schlegel \etal\ 1998}\nocite{schlegel98}) have
apparently conspired to keep this nearby system away from detailed
observational scrutiny to date.  Below we discuss the first spatially
resolved study of the ISM and stars within the Orion dwarf; we find
modest-level SF in a dark-matter dominated disk that harbors a wealth
of small-scale structure.

The distance of the Orion dwarf remains uncertain, primarily due to
its low galactic latitude and the resulting difficulty of resolved
stellar population work.  Previous authors have suggested distances
based on the brightest stars method after solving for foreground
extinction; for example, \citet{karachentsev96} find D = 6.4 Mpc.
More recent infrared work by \citet{vaduvescu05} lowers this value to
5.4 Mpc; we adopt this latter value in the present work, but note that
the distance remains a significant source of uncertainty.  We
summarize the basic parameters of the Orion dwarf in Table~\ref{t1}.

\section{Observations and Data Reduction}
\label{S2}

\subsection{{\it KPNO 0.9m} Imaging}
\label{S2.1}

New optical images of the Orion dwarf were acquired with the {\it KPNO
0.9m} telescope\footnote{The {\it KPNO} 0.9-meter telescope is
operated jointly by a consortium that includes the University of
Florida, Indiana University, San Francisco State University,
Wesleyan University, the University of Wisconsin-Madison, the
University of Wisconsin-Oshkosh, the University of Wisconsin-Stevens
Point, and the University of Wisconsin-Whitewater}.  Two 20-minute
exposures in a narrowband \halpha\ filter, and one 4-minute exposure
in a broadband R filter, were acquired on 11 March, 2008.
Figure~\ref{figcap1} presents the R-band and continuum-subtracted
\halpha\ images, and compares these with a near-infrared \ks-band
image kindly provided by O. Vaduvescu.

The broadband images reveal a smooth stellar component, extending from
northeast to southwest, that is $\sim$300\arcsec\ ($\sim$7.8 kpc at
5.4 Mpc) in diameter.  While the stellar density distribution results
in an overall low-surface brightness appearance, the central stellar
component is reasonably symmetric and well-defined.  We will refer to
this $\sim$8 kpc diameter stellar distribution as the central stellar
component throughout the rest of the paper.  The high foreground
contamination from Milky Way stars is clear in the broadband imges.
Most of the bright point sources within the disk are foreground stars;
the continuum subtraction process (see below) shows that only a few
are stellar clusters associated with ongoing SF in the Orion dwarf
itself.  The most prominent of these clusters are discussed further
below.

We use standard techniques to subtract the continuum from the
narrowband \halpha\ images.  First, the images are cosmic ray
rejected, aligned and combined.  The broad and narrowband images are
then smoothed to a common point spread function size and flux scaled
to remove the continuum from the narrowband image.  Finally, standard
photometry routines are applied, using observations of photometric
standard stars acquired during the same observing session.  We
estimate the photometric accuracy of our final images to be 10\% or
better.

Comparing the continuum-subtracted \halpha\ image with the broadband R
and \ks-images in Figure~\ref{figcap1}, we find that most of the
active SF probed by our \halpha\ imaging is contained in moderate-flux
\HII\ regions scattered widely throughout the disk.  We catalog 18
individual \HII\ regions, many of which are located well outside of
the central stellar component as seen in the R and \ks-band images.
Only three regions (labeled as \#3, 10, 11 in Figure~\ref{figcap1})
are associated with high surface brightness stellar clusters in the
broadband images.  We catalog the positions, observed and (Galactic)
extinction-corrected fluxes of these \HII\ regions in Table~\ref{t2}.
Note that \HII\ region \#11 consists of three distinct peaks that are
just barely resolved; the flux in the table is the sum of these three.

From the continuum-subtracted \halpha\ image we derive a total
\halpha\ flux (corrected for foreground extinction) of
(113\,$\pm$\,1.3)\,$\times$\,10$^{-14}$ \ergseccm.  At the adopted
distance of 5.4 Mpc, this corresponds to a luminosity of
4.0\,$\times$\,10$^{39}$ erg\,s$^{-1}$.  The Orion dwarf is thus a
comparatively quiescent and, as shown in the following sections,
gas-rich, dwarf galaxy.  We note with interest that the \halpha\
luminosity is close to the transition value found by \citet{lee09}
where \halpha\ emission begins to systematically under-predict the
current star formation rate in dwarf galaxies.  Using their
prescription (which was constructed using a combination of UV and
\halpha\ indicators), we derive a current star formation rate of
$\sim$0.04 \msun\,yr$^{-1}$; this value is roughly 30\% higher than
the value found by a straightforward application of the
\citet{kennicutt98a} prescription.  While the uncertainties are
substantial (foreground extinction, adopted distance), the Orion dwarf
appears to be in a state of relative quiescence at the present time.

\subsection{{\it VLA} Spectral Line Observations}
\label{S2.2}

\HI\ spectral line imaging of the Orion dwarf galaxy was obtained with
the {\it Very Large Array} ({\it VLA}\footnote{The National Radio
Astronomy Observatory is a facility of the National Science Foundation
operated under cooperative agreement by Associated Universities,
Inc.}) on 8 March, 2008.  These data were acquired under the auspices
of the {\it VLA/VLBA Observing for University Classes} program.  The
array was in the C configuration during these observations; the total
on-source integration time was 140 minutes.  We supplemented these
data with dynamically-scheduled observations with the array in the B
configuration on 12 January, 2008 (total on-source integration time
$=$ 127 minutes).  A total bandwidth of 3.125 MHz was used, with 128
channels separated by 24.4 kHz (5.12 \kms).  All reductions were
performed using the Astronomical Image Processing System ({\sc AIPS})
package.

Data handling followed standard procedures, with necessary
modifications to account for the array being in a ``hybrid'' state
containing both {\it VLA} and {\it EVLA} receivers.  12 of 27 antennae
were fitted with {\it EVLA} receivers during the B configuration
observations; in the C configuration there were 13 {\it EVLA}-equipped
antennae.  While a complete discussion of the intricacies of data
handling during the {\it VLA} to {\it EVLA} transition is beyond the
scope of this paper \footnote{We refer the reader to www.nrao.edu for
detailed discussion.}, we briefly note the most significant changes to
the ``standard'' spectral line reduction steps here.  First, an
alternative channel zero dataset was created that averaged together
central channels well separated from the edges of the bandpass.
Second, each calibration step is bracketed by a flag/unflag step for
all EVLA-EVLA baselines.  The combination of these steps allows us to
retain the maximum number of baselines from our observations.

Other than the modifications above (and discussed further on the {\it
VLA} website), the data handling proceeded normally.  First,
interference and bad data were removed from each dataset; flux, gain
and phase calibrations were applied, derived from observations of
0542$+$498 (3C147; primary calibrator) and 0532$+$075 (secondary
calibrator) acquired during each observing session.  Based on the
measured flux density of the phase calibrator from each observing
session, we estimate our absolute flux calibration to be accurate at
the $\sim$15\% level.  The calibrated {\it u-v} datasets were then
combined and the underlying continuum was removed.

These combined datasets were then imaged and cleaned using the {{\sc
IMAGR} task in {\sc AIPS}.  We produce two datacubes, convolved to
circular beam sizes of 10\arcsec\ and 20\arcsec.  The rms noise
levels per 5.2 \kms\ channel are 1.51 and 0.55 K for the 10\arcsec\
and 20\arcsec\ cubes, respectively; the corresponding 3\,$\sigma$ \HI\
column density sensitivities are N$_{\rm
HI}$=(4.3,\,1.5)$\times$10$^{19}$\,cm$^{-2}$, respectively.  We
explicitly account for the different beam sizes of the dirty and clean
maps using residual flux scaling (e.g., {J\"ors\"ater \& van Moorsel
1995}\nocite{jorsater95}; {Walter \etal\ 2007}\nocite{walter07}).

Moment maps at 10\arcsec\ and 20\arcsec\ resolution (representing
integrated intensity, velocity, and velocity dispersion) were created
from these data cubes using an approach similar to the one described
and applied in \citet{cannon09}.  First, the ``low-resolution''
(20\arcsec\ beam) cube was blanked at the 2\,$\sigma$ level.  Channels
containing line emission were then inspected individually; real and
spurious emission were differentiated by requiring structures to be
present in 3 or more consecutive channels.  This edited cube is used
to blank both the ``low'' and ``high-resolution'' cubes; this assures
that the same regions contribute to both final moment maps.  The
images of integrated \HI\ emission are discussed in detail in
\S~\ref{S3}.

The total \HI\ flux integral, derived from the 20\arcsec\ cube, is
50.3\,$\pm$\,5.1 Jy\,\kms; this is lower than the single-dish flux
density of 80.59$\pm$7.72 Jy\,\kms\ found by {Springob \etal\
(2005}\nocite{springob05}; note that those authors apply a $\sim$8\%
correction for \HI\ self-absorption).  This difference likely arises
from missing short spacings that provide sensitivity to diffuse
structure; further observations with the array in the compact D
configuration would be fruitful in this regard.  Using these numbers
we derive a total \HI\ mass of (3.5$\pm$0.5)\,$\times$\,10$^{8}$
\msun; including a 35\% correction for Helium and molecular material,
we adopt (4.7$\pm$0.7)\,$\times$\,10$^{8}$ \msun\ as the total gas
mass (M$_{\rm gas}$) of the Orion dwarf.

During the reduction of the \HI\ spectral line data, we averaged 78
line-free channels to produce a 1.4 GHz radio continuum image.  This
image has a beam size of 13.8\arcsec\,$\times$\,12.9\arcsec\ and an
RMS noise level of 0.17 \mjypbm. We do not detect radio continuum
emission from the Orion dwarf at this sensitivity level.  The
5$\sigma$ upper limit to the global 1.4 GHz radio continuum flux
density is derived to be S$_{\rm 1.4 GHz} \lsim$0.85 mJy.  This is
consistent with the NVSS non-detection \citep{condon92}.  The expected
thermal radio continuum flux density, based on the (foreground
extinction-corrected) \halpha\ flux, is between 0.9--2.0 mJy
(depending on the adopted electron temperature; see {Caplan \&
Deharveng 1986}\nocite{caplan86}).  Examining the continuum emission
from the individual \HII\ regions also shows they they are beyond the
sensitivity of our L-band continuum image.  The most luminous \HII\
region (region \#11) has an \halpha\ flux that would produce a thermal
radio continuum flux of 0.28 mJy (assuming T$_{\rm e}$ = 10$^4$ K).
This is again well below our detection threshold.  Thus, even if
nonthermal processes dominate the radio continuum flux in the Orion
dwarf, it is below our current sensitivity level to study in detail.

\section{Gaseous, Stellar, and Dark Components}
\label{S3}
\subsection{The Neutral Gas Distribution and Dynamics}
\label{S3.1}

\HI\ emission from the Orion dwarf is detected in 28 channels of our
final datacubes, spanning the heliocentric radial velocity range of
301.8 -- 441.0 \kms.  Figure~\ref{figcap2} shows the individual
channel maps of the 20\arcsec-resolution datacube.  The ordered
rotation of the system is prominent, as is a wealth of structure that
permeates the ISM.  Especially evident is a kinematic break in the
distribution and velocity structure of the gas in the central regions
of the disk (consider the panels of Figure~\ref{figcap2} surrounding
379.2 \kms).  We explore this structure in more detail below.

Integrating the sum of the flux in each of the channel maps shown in
Figure~\ref{figcap2} allows us to extract the global \HI\ profile
shown in Figure~\ref{figcap3}.  As expected from the rotation of the
system evident in Figure~\ref{figcap2}, we see a very symmetric,
double-peaked \HI\ profile.  We derive an \HI\ systemic velocity of
368.5\,\pom\,1.0 \kms; this velocity is noted in Figure~\ref{figcap3}.
The full width of the profile at half maximum is 118\,\pom\,4 \kms;
this can be compared with line widths between 97 and 129 \kms\ in
{Springob \etal\ (2005}\nocite{springob05}; the different techniques
used to extract the line widths in that work account for the range of
derived values).

In Figure~\ref{figcap4} we present \HI\ moment zero (integrated
intensity) images of the Orion dwarf at 10\arcsec\ and 20\arcsec\
resolutions. As noted for the individual channel maps, the \HI\ disk
shows rich structure at these physical resolutions.  The outer disk
contains tenuous \HI\ gas, but column densities rise above the
5\,$\times$\,10$^{20}$ cm$^{-2}$ level at intermediate radii.  There
is plentiful high-column density ($>$10$^{21}$ cm$^{-2}$) \HI\
throughout the disk.  It is interesting, however, that the inner
regions of the \HI\ disk are not the regions with the highest gas
surface densities.  The kinematic features noted in the channel maps
are apparently associated with prominent regions of low column density
($<$3\,$\times$\,10$^{20}$ cm$^{-2}$) in the inner regions.  We
hereafter refer to these regions as \HI\ ``depressions'' or ``holes''
(see further discussion below).

The \HI\ disk is significantly larger than the central stellar
component of the system; Figure~\ref{figcap5} compares the optical
R-band and \halpha\ images with the \HI\ distribution (shown by the
10$^{21}$ cm$^{-2}$ contour).  Note that most of the the central
stellar component (traced by the R-band image) is located near the
\HI\ depressions.  In contrast, most (but not all) of the high surface
density \HI\ gas (N$_{\rm HI}$ $>$ 10$^{21}$ cm$^{-2}$) is exterior to
the red stellar population.

Figure~\ref{figcap5} also compares the locations of high-column
density \HI\ gas and high surface brightness \HII\ regions. We find a
strong correlation between \HI\ gas at or above the 10$^{21}$
cm$^{-2}$ level and regions of active SF as traced by \halpha\
emission.  This is further verification of the well-known empirical
``star formation threshold'' (e.g., {Skillman
1987}\nocite{skillman87}, {Kennicutt 1989}\nocite{kennicutt89},
{Kennicutt 1998b}\nocite{kennicutt98b}, and references therein): high
\HI\ column densities are often co-spatial with regions of recent SF.
We note that only two of our 18 detected \HII\ regions (specifically,
regions 7 and 9 in Table~\ref{t2}; see also Figure~\ref{figcap5}) are
separated by one 10\arcsec\ beam element or more from \HI\ column
densities of 10$^{21}$ cm$^{-2}$.

The channel maps shown in Figure~\ref{figcap2} suggest well-ordered
rotation throughout the disk.  The first moment image (representing
intensity-weighted velocity) presented in Figure~\ref{figcap6}
confirms this.  Isovelocity contours are symmetric in the outer disk,
indicative of solid-body rotation.  In the central regions of the
disk, however, the \HI\ ``holes/depressions'' manifest a pronounced
kink in these contours (consider the contours at 370\,\pom\,20 \kms).
This suggests a significant departure from ordered rotation (i.e.,
non-circular motion) within the inner disk of the Orion dwarf.

In Figure~\ref{figcap7} we compare the \HI\ column density
distribution with the second moment image (representing intensity
weighted velocity dispersion).  Throughout most of the disk, the
dispersion is nearly constant at $\sigma_{\rm V}$ $\simeq$ 7\,$\pm$\,2
\kms.  However, this dispersion increases markedly in the inner disk
region, corresponding roughly to the location of the central stellar
component.  There, values reach $\sim$15 and 20 \kms\ in the
20\arcsec\ and 10\arcsec\ resolution maps, respectively.  While beam
smearing effects add uncertainty, we can conclude with confidence that
the ``depressions/holes'' in the inner ISM correlate with areas of
higher than average velocity dispersion.

The proximity of this galaxy, and the well-ordered rotation throughout
most of the outer \HI\ disk, are conducive to rotation curve analysis.
To this end, we fitted tilted ring models to the velocity field using
tools in the {\it GIPSY} software package\footnote{The Groningen Image
Processing System (GIPSY) is distributed by the Kapteyn Atronomical
Institute, Groningen, Netherlands.}.  In an iterative sequence, we fit
the systemic velocity (V$_{\rm sys}$), dynamical center position,
position angle (P.A.), inclination ({\it i}) and rotational velocity
for the galaxy as a whole, and for the receding and approaching sides
individually.  Optimal fits were obtained with {\it i} =
(45$\pm$3)\arcdeg, P.A. = (20$\pm$2)\arcdeg, V$_{\rm sys}$ =
(368.5$\pm$1.0) \kms, and a dynamical center at ($\alpha$,$\delta$,
J2000) = (05:45:01.66, 05:03:55.2).  The resulting curves, shown at
both 10\arcsec\ and 20\arcsec\ resolution in Figure~\ref{figcap8},
demonstrate solid-body rotation to $\sim$140\arcsec\ (3.7 kpc) and a
flattening at $\sim$82 \kms\ to the detection limit in the outer disk
($\sim$6.8 kpc).  Assuming circular orbits, the implied dynamical mass
at this radius is 10.6$\times$10$^{9}$ \msun.

Our tilted ring analysis also allows us to examine the radial behavior
of \HI\ mass density per unit area throughout the gas disk.  We again
used the {\it GIPSY} software package to integrate the \HI\ flux per
unit area, in concentric rings separated by the beam widths, after
correcting for inclination and the galaxy's major axis position angle.
The resulting plot, shown in Figure~\ref{figcap9}, clearly
demonstrates that the inner region of the \HI\ disk where the
``depressions/holes'' are located contains significantly less neutral
gas per unit area than the regions further out in the disk.  The
highest mass surface densities are located between $\sim$1.5--3 kpc
from the dynamical center; the inner disk has a mass surface density
$\sim$ 40\% lower at these physical resolutions.

\subsection{The Old and Young Stellar Populations}
\label{S3.2}

As noted above, the dynamical mass exceeds the gaseous mass (even when
accounting for missing \HI\ flux compared to single-dish
observations).  We now explore what fraction of this dynamical mass is
in the form of stars.  The underlying stellar mass is estimated using
the near-infrared (IR) photometry (J and \ks\ bands) presented by
\citet{vaduvescu05}.  Those authors find (J\,$-$\,\ks) $=$ $+$0.80 and
a total \ks\ magnitude of $+$10.90. When comparing to models (see
below) we assume that the color difference between K and \ks\ is
negligible; further, we assume L$_{\rm K,\,\odot} = +$3.33
\citep{cox00,bessel79}.  Accounting for extinction (see above), the
total K-band luminosity of the Orion dwarf is
$\sim$3.5\,$\times$\,10$^8$ L$_{\odot}$.

To estimate the mass of the stellar component, we apply the models
presented in \citet{bell01}.  Using an observed color, one can infer a
stellar mass-to-light ratio (M/L) for a variety of stellar population
synthesis models that cover a range of metallicities, initial mass
functions, and galaxy evolution properties.  The near-infrared colors
are least susceptible to variations due to internal extinction or
recent SF.  In order to compare the near-IR photometry with the
\citet{bell01} models, we followed the approach discussed in
\citet{lee06} and derived a general relation between (J\,$-$\,K) color
and the stellar M/L ratio as a linear combination of (V\,$-$\,J) and
(V\,$-$\,K) indices.  Using the sub-solar metallicity models in
\citet{bell01}, this yields estimates of the stellar M/L ratio for
each model.  Using these M/L ratios and the observed K-band
luminosity, we then obtain estimates of the galaxy's stellar mass.
Using the dispersion amongst the low-metallicity models as an
indicator of the accuracy of this method, we conclude that the
underlying stellar mass in the Orion dwarf is
(3.7\,$\pm$\,1.5)\,$\times$\,10$^{8}$ \msun.  A similar calculation
that uses all available models in the \citet{bell01} work (i.e.,
including those for all metallicities) finds a similar (equal within
errors), but slightly lower, value of
(2.9\,$\pm$\,1.3)\,$\times$\,10$^{8}$ \msun.

This stellar mass is comparable to, but on the high end of, the
distribution of stellar masses in local dwarfs found by \citet{lee06}
While that investigation uses a longer color baseline, our techniques
for deriving the stellar masses are identical.  The Orion dwarf has a
stellar mass comparable to that of the well-studied irregulars
NGC\,4214 and NGC\,55; its current (\halpha-derived) star formation
rate is lower, however, than in those systems.

\subsection{A Dark Matter Dominated Galaxy}
\label{S3.3}

Like most dwarf galaxies, the Orion dwarf appears to be dark-matter
dominated throughout the disk.  In \S~\ref{S2.2} we derived a total
gas mass M$_{\rm gas}$ = (4.7$\pm$0.7)\,$\times$\,10$^{8}$ \msun; this
includes a 35\% correction for Helium and molecular material.  From
\S~\ref{S3.2}, the total stellar mass M$_{\star}$ =
(3.7\,$\pm$\,1.5)\,$\times$\,10$^{8}$ \msun.  The resulting mass ratio
of gas to stars, log(M$_{\rm gas}$/M$_{\rm \star}$) $\simeq$ 0.10, is
comparable to many systems in the \citet{lee06} study.

The masses of these luminous components can be compared with the total
dynamical mass of the system (from \S~\ref{S3.1}, M$_{\rm dyn}$
$\simeq$ 10.6$\times$10$^{9}$ \msun).  This implies a M/L ratio of
$\sim$13 for the galaxy as a whole.  Continuing inward through the
disk the M/L ratio remains above unity; most of the \ks-band
luminosity is contained within $\sim$3\arcmin\ of the dynamical center
(see, for example, Figures~\ref{figcap1} and \ref{figcap5}). At the
point where the rotation curve flattens ($\simeq$140\arcsec\ = 3.4
kpc) the rotation velocity is $\sim$79 \kms; the interior dynamical
mass is $\sim$5.3$\times$10$^{9}$ \msun, already exceeding the total
luminous mass in the entire galaxy.

\section{The Star Formation - ISM Connection}
\label{S4}

The rich ISM structure seen in the integrated \HI\ distribution (see
Figure~\ref{figcap4}), and the locations of the stellar distribution
and the sites of recent SF within the disk (see Figure~\ref{figcap5}),
are suggestive of a formative link between recent SF and the observed
structures of the ISM. As noted in \S~\ref{S3.1}, the most prominent
kinematic features are apparent in both the individual channel maps
(see Figure~\ref{figcap2}) and in the integrated distribution (see
Figure~\ref{figcap5}) as regions of low column density.  We now
explore the nature of the three most prominent of these
``depressions/holes'' in more detail.

To better elucidate the positions and morphologies of these features,
we overlay their positions on the \HI\ integrated intensity map, the
\HI\ velocity field, and the \HI\ velocity dispersion map in
Figure~\ref{figcap10}.  We number these features in order of
increasing right ascension and summarize their basic properties in
Table~\ref{t3}.  The ``depressions/holes'' are spatially coincident
with departures from ordered isovelocity contours, have lower than
average \HI\ column density, and have higher than average \HI\
velocity dispersion.

Figure~\ref{figcap11} shows these same ellipses overlaid on optical
R-band and continuum-subtracted \halpha\ images; the field of view is
the same as in Figure~\ref{figcap5}.  Each of the kinematic features
is completely enclosed within the N$_{\rm HI}$ = 10$^{21}$ cm$^{-2}$
contour.  None of them is cospatial with the central stellar disk;
each is offset by some amount, with feature \#3 showing the least
positional agreement with the stellar distribution.

When comparing the locations of these features with the individual
channel maps, however, ``depression/hole'' \#3 (which shows the
poorest positional agreement with the stellar distribution) is the
easiest to identify.  Examining Figure~\ref{figcap12}, which shows the
same channel maps as Figure~\ref{figcap2} but now overlaid with the
locations of the ``depressions/holes'', this feature is clearly
defined as a dearth of \HI\ emission in the panels near 379 \kms.
As expected, this feature causes non-uniformities in the integrated
velocity field; Figure~\ref{figcap10} shows this quite clearly.

If these features are in fact holes in the ISM, it is natural to
search for kinematic verification of their expansion.  To this end, we
searched the datacubes for kinematic signatures of expansion of the
major observed \HI\ structures.  We examined both position-velocity
and radius-velocity diagrams throughout the \HI\ disk.  These features
are each detected as pronounced breaks in the local \HI\ distribution
as seen in position-velocity space.  We show one such profile in
Figure~\ref{figcap13}; in both major- and minor-axis cuts, the
``depression/hole'' features are clearly evident as kinematic
discontinuities in the inner disk.  

While the position-velocity diagrams verify the lower-than-average
column densities of these features, radius-velocity diagrams do not
find unambiguous evidence for expansion of these structures at the
present time.  If these features are expanding within the disk, they
must be doing so at a modest rate (V$_{\rm exp} \lsim$ 10 \kms, or
roughly two channel widths).  We note that significant expansion may
go undetected within these data if that motion is significantly offset
with respect to the galaxy's disk.

These types of features in the ISM of nearby galaxies are often
attributed to stellar evolution processes (i.e., stellar winds and
supernovae).  In the simplest sense, these processes deposit energy
into the ISM of the host galaxy; this energy evacuates the gas
surrounding regions of strong recent SF.  The ``simplistic'' nature of
dwarfs is advantageous in this regard, as such features may be
comparatively long-lived.  This ``feedback'' mechanism thus provides a
direct link between the stellar and gaseous components of a galaxy.

Recent work by \citet{weisz09} has suggested that stellar evolution
processes are energetically capable of producing adequate energies to
drive expanding shells in the ISM of low-mass galaxies.  An important
step toward understanding this mechanism was taken by other works that
suggest that these stellar evolution processes need not necessarily be
``instantaneous''.  Rather, recent observations have revealed that
periods of elevated star formation rates can be sustained in low-mass
galaxies for hundreds of Myr (see {Cannon \etal\
2003}\nocite{cannon03}, {Weisz \etal\ 2008}\nocite{weisz08}, {Mc~Quinn
\etal\ 2009}\nocite{mcquinn09}).  Such events can inject large
energies into the ISM without the requisite of leaving a remnant
stellar cluster (signifying an instantaneous burst) behind.

Unfortunately we are unable to directly test this model with our
present \HI\ imaging of the Orion dwarf.  An estimate of the energetic
requirements to evacuate these features is very difficult without a
measurement of the present expansion rate.  To accurately infer the
energetic requirements, one would ideally model the underlying mass
distribution using deep infrared imaging \citep{oh08} and use this, in
combination with the \HI\ velocity dispersion, to derive the \HI\
scale height as a function of radius.  These values can then be used
to estimate the \HI\ number density in the gas prior to the creation
of the structure.  The energetic requirement is then straightforward:
move a known quantity of gas by a measured amount in a given amount of
time (based on the ``kinematic age'' of the structure, which depends
on its expansion velocity).  To draw definitive conclusions on the
``feedback'' scenario, one also must have deep, single-star photometry
in order to reconstruct the recent star formation history (see {Weisz
\etal\ 2008}\nocite{weisz08} and {2009}\nocite{weisz09}).
Unfortunately, the large foreground extinction toward the Orion dwarf
makes such an endeavor very challenging.

If the observed structures are in fact the products of stellar
evolution, then, adopting the formalism of \citet{brinks86} and
\citet{walter99}, these structures may be classified as regions of
``total blowout''.  Any hole or shell structures have either stalled
completely (meaning they have reached pressure equilibrium with the
surrounding ISM or have broken out perpendicular to the disk) or were
sufficiently extended (spatially or temporally) so as to preclude the
formation of a single observable kinematic feature.  Such a scenario
appears increasingly reasonable based on recent studies of the star
formation histories of low-mass galaxies ({Weisz \etal\
2008}\nocite{weisz08} and {2009}\nocite{weisz09}, {Mc~Quinn \etal\
2009}\nocite{mcquinn09}), but we must await further data to study this
hypothesis in the Orion dwarf.

We note that various alternatives to the stellar evolution model have
been postulated (see, for example, the discussion in {Rhode \etal\
1999}\nocite{rhode99}).  Given the number of assumptions required to
derive the creation energy of an expanding structure from
observations, it is not surprising that one of the leading
alternatives is simply that the energetics are incorrectly attained.
We note that the methods for deriving these numbers are using
increasing levels of sophistication (e.g., {Oh \etal\
2008}\nocite{oh08}).  Similarly, spatially resolved stellar population
work is helping to eliminate non-standard initial mass functions as
the explanation.  Other hypotheses are more challenging to disregard
and they remain theoretically viable.  Some of these include high
velocity cloud impacts, disk instabilities, turbulence, or ram
pressure stripping (e.g., {S{\' a}nchez-Salcedo
2002}\nocite{sanchezsalcedo02}).

\section{Conclusions}
\label{S5}

New \HI\ and optical imaging of the Orion dwarf have been presented.
The stellar component occupies the inner region of a mostly
well-ordered \HI\ disk.  Column densities rise above the 10$^{21}$
cm$^{-2}$ level in multiple areas, although the central stellar
component is primarily coincident with lower surface density gas in
the inner disk.  This area contains three regions of lower than
average column density, higher than average velocity dispersion, and
departures from symmetric isovelocity contours; we term these features
``holes/depressions''.  The areas of active SF as traced by \halpha\
emission show a strong preference for high surface density gas; some
of the active regions are well outside of the central stellar
component of the system.

The Orion dwarf is well-suited for rotation curve analysis.  Through
an iterative procedure we extract a rotation curve that is
essentially flat at $\sim$82 \kms\ out to the edge of the \HI\ disk
($\simeq$6.8 kpc).  At the last rotation curve point, the implied
dynamical mass M$_{\rm dyn}$ $\simeq$ 10.6$\times$10$^{9}$ \msun.
Applying the models of \citet{bell01}, we use optical and infrared
images to derive a stellar mass of
(3.7\,$\pm$\,1.5)\,$\times$\,10$^{8}$ \msun.  This can be compared
with the neutral gas mass of (4.7\,$\pm$\,1.5)\,$\times$\,10$^{8}$
\msun\ (which includes a correction for helium and molecular
material).  The Orion dwarf is thus dark matter dominated throughout
the disk.

We investigate the kinematics of the ``hole/depression'' regions.
While these appear as prominent breaks in position-velocity space, and
are evident in individual channel maps, a radius-velocity analysis
does not show signs of expansion of these features at the present
sensitivity and velocity resolution.  Unfortunately, this lack of a
measured expansion velocity for these features precludes a direct 
derivation of the energetic requirements for their formation.

While the evidence for ``feedback'' (i.e., the effects of energy
released by stellar evolution processes on the surrounding ISM) is
tantalizing, we are unable to directly test this hypothesis.  In
addition to higher velocity resolution \HI\ imaging (necessary to
measure a potentially slower expansion rate for the
``hole/depression'' structures), we also require deep, single-star
photometry to unambiguously connect the past SF with the current ISM
features.  The high foreground reddening toward the Orion dwarf will
make this challenging; however, the curious properties of this system
suggest that the investment will be a worthwhile one in the quest to
connect SF activity with characteristics of the ISM in dwarf galaxies.

\acknowledgements

The authors would like to thank the NRAO for making the {\it VLA/VLBA
Observing for University Classes} Program available to the educational
community, and for travel support and hospitality while at the Array
Operations Center.  We would also like to thank Robert Dickman, Vivek
Dhawan, and the AOC staff for organizing a most productive and
enjoyable visit.  Thanks to Ovidiu Vaduvescu for allowing us to
analyze and present the infrared image in this manuscript.  We
acknowledge a helpful anonymous referee whose comments improved this
manuscript.  JMC thanks Macalester College for research and teaching
support that made this project possible, and thanks Henry Lee for
helpful discussions.  The optical observations were made possible
through financial support from Indiana University and a grant from the
National Science Foundation (AST-0823801) to JJS.  This research has
made use of the NASA/IPAC Extragalactic Database (NED) which is
operated by the Jet Propulsion Laboratory, California Institute of
Technology, under contract with the National Aeronautics and Space
Administration, and NASA's Astrophysics Data System.

\clearpage
\bibliographystyle{apj}                                                 


\clearpage
\begin{deluxetable}{lcc}  
\tablecaption{Basic Characteristics of the Orion Dwarf Galaxy} 
\tablewidth{0pt}  
\tablehead{ 
\colhead{Parameter} &\colhead{Value}}    
\startdata      
Right ascension (J2000)          &05$^{\rm h}$ 45$^{\rm m}$ 01.$^{\rm s}$6\\        
Declination (J2000)              &+05\arcdeg 03\arcmin 41\arcsec\\    
Adopted distance (Mpc)                            &5.4\tablenotemark{a}\\
A$_{\rm R}$                      &1.959\tablenotemark{b}\\
Interferometric S$_{\rm HI}$ (Jy km/s)           &50.3\,\pom\,5.1\\
Single-dish S$_{\rm HI}$ (Jy km/s)          &80.59\,\pom\,7.72\tablenotemark{c}\\
\HI\ mass M$_{\rm HI}$ (\msun)   &(3.46 \pom 0.5)\,$\times$\,10$^8$\\
\enddata     
\label{t1}
\tablenotetext{a}{\citet{vaduvescu05}}
\tablenotetext{b}{\citet{schlegel98}}
\tablenotetext{c}{\citet{springob05}}
\end{deluxetable}   

\clearpage
\begin{deluxetable}{lcccc}
\tablecaption{\halpha\ Emission Line Region Catalog}
\tablewidth{0pt}  
\tablehead{ 
\colhead{Number}    &\colhead{$\alpha$\tablenotemark{a}} &\colhead{$\delta$\tablenotemark{a}} &\colhead{Observed \halpha\ Flux} &\colhead{Extinction-Corrected \halpha\ Flux\tablenotemark{b}}\\
    &\colhead{(J2000)} &\colhead{(J2000)} &\colhead{(10$^{-14}$ erg\,s$^{-1}$\,cm$^{-2}$)} &\colhead{(10$^{-14}$ erg\,s$^{-1}$\,cm$^{-2}$)}}
\startdata      
1   &05$^{\rm h}$\,44$^{\rm m}$\,56.47$^{\rm s}$ &05\arcdeg\,02\arcmin\,44.44\arcsec    &0.57\pom0.06 &3.46\pom0.36\\  
2   &05$^{\rm h}$\,44$^{\rm m}$\,57.95$^{\rm s}$ &05\arcdeg\,03\arcmin\,20.11\arcsec    &0.64\pom0.04 &3.89\pom0.24\\  
3   &05$^{\rm h}$\,44$^{\rm m}$\,59.70$^{\rm s}$ &05\arcdeg\,04\arcmin\,03.90\arcsec    &0.98\pom0.04 &5.95\pom0.24\\  
4   &05$^{\rm h}$\,44$^{\rm m}$\,59.99$^{\rm s}$ &05\arcdeg\,03\arcmin\,25.04\arcsec    &2.05\pom0.06 &12.45\pom0.36\\  
5   &05$^{\rm h}$\,45$^{\rm m}$\,00.39$^{\rm s}$ &05\arcdeg\,03\arcmin\,03.75\arcsec    &1.24\pom0.05 &7.53\pom0.30\\  
6   &05$^{\rm h}$\,45$^{\rm m}$\,00.44$^{\rm s}$ &05\arcdeg\,03\arcmin\,15.33\arcsec    &0.60\pom0.04 &3.64\pom0.24\\ 
7   &05$^{\rm h}$\,45$^{\rm m}$\,00.87$^{\rm s}$ &05\arcdeg\,05\arcmin\,38.32\arcsec    &0.23\pom0.03 &1.40\pom0.18\\  
8   &05$^{\rm h}$\,45$^{\rm m}$\,00.99$^{\rm s}$ &05\arcdeg\,03\arcmin\,15.88\arcsec    &1.01\pom0.05 &6.13\pom0.30\\ 
9   &05$^{\rm h}$\,45$^{\rm m}$\,01.68$^{\rm s}$ &05\arcdeg\,02\arcmin\,44.11\arcsec    &0.82\pom0.04 &4.98\pom0.24\\   
10  &05$^{\rm h}$\,45$^{\rm m}$\,01.94$^{\rm s}$ &05\arcdeg\,04\arcmin\,07.87\arcsec    &2.25\pom0.06 &13.66\pom0.36\\  
11  &05$^{\rm h}$\,45$^{\rm m}$\,02.61$^{\rm s}$ &05\arcdeg\,04\arcmin\,19.13\arcsec    &3.92\pom0.09 &23.81\pom0.55\\  
12  &05$^{\rm h}$\,45$^{\rm m}$\,02.69$^{\rm s}$ &05\arcdeg\,02\arcmin\,40.95\arcsec    &0.68\pom0.06 &4.13\pom0.36\\  
13  &05$^{\rm h}$\,45$^{\rm m}$\,03.11$^{\rm s}$ &05\arcdeg\,04\arcmin\,39.36\arcsec    &0.66\pom0.04 &4.01\pom0.24\\  
14  &05$^{\rm h}$\,45$^{\rm m}$\,03.21$^{\rm s}$ &05\arcdeg\,05\arcmin\,16.82\arcsec    &0.25\pom0.03 &1.52\pom0.18\\  
15  &05$^{\rm h}$\,45$^{\rm m}$\,03.95$^{\rm s}$ &05\arcdeg\,02\arcmin\,41.47\arcsec    &0.30\pom0.05 &1.82\pom0.30\\  
16  &05$^{\rm h}$\,45$^{\rm m}$\,04.26$^{\rm s}$ &05\arcdeg\,04\arcmin\,48.27\arcsec    &0.89\pom0.05 &5.41\pom0.30\\       
17  &05$^{\rm h}$\,45$^{\rm m}$\,04.59$^{\rm s}$ &05\arcdeg\,05\arcmin\,02.30\arcsec    &0.67\pom0.03 &4.07\pom0.18\\  
18  &05$^{\rm h}$\,45$^{\rm m}$\,05.53$^{\rm s}$ &05\arcdeg\,03\arcmin\,25.42\arcsec    &0.91\pom0.06 &5.53\pom0.36\\ 
\enddata
\label{t2}
\tablenotetext{a}{The right ascension and declination at the estimated central position of
the \halpha\ region; estimated astrometric uncertainty is
\pom0.5\arcsec.}
\tablenotetext{b}{The \halpha\ fluxes are corrected for the R-band extinction level of 1.959 magnitudes
({Schlegel \etal\ 1998}\nocite{schlegel98}).} 
\end{deluxetable}  

\clearpage
\begin{deluxetable}{lcccccccc}
\tabletypesize{\scriptsize}
\tablecaption{Prominent \HI\ ``Holes/Depressions'' in the Orion Dwarf Galaxy} 
\tablewidth{0pt}  
\tablehead{ 
\colhead{Number}    &\colhead{$\alpha$}                 &\colhead{$\delta$}                 &\colhead{Major Axis}            &\colhead{Minor Axis}            &\colhead{Position Angle}\\
\colhead{}          &\colhead{(J2000)\tablenotemark{a}} &\colhead{(J2000)\tablenotemark{a}} &\colhead{(pc)\tablenotemark{b}} &\colhead{(pc)\tablenotemark{b}} &\colhead{(degrees)\tablenotemark{c}}}
\startdata      
\vspace{0.0 cm}      
1   &05$^{\rm h}$\,45$^{\rm m}$\,00.76$^{\rm s}$ &05\arcdeg\,04\arcmin\,30.5\arcsec    &920  &650 &160\\  
2   &05$^{\rm h}$\,45$^{\rm m}$\,02.47$^{\rm s}$ &05\arcdeg\,03\arcmin\,19.0\arcsec    &1440 &770 &150\\  
3   &05$^{\rm h}$\,45$^{\rm m}$\,04.45$^{\rm s}$ &05\arcdeg\,04\arcmin\,05.0\arcsec    &1700 &920 &40\\  
\enddata     
\label{t3}
\tablenotetext{a}{The right ascension and declination of the center of
the structure; estimated astrometric uncertainty is \pom3\arcsec.  The locations are shown 
in panels {\it c} and {\it d} of Figure~\ref{figcap5}.}
\tablenotetext{b}{The major and minor axis diameters in pc, using the assumed distance 
of 5.4 Mpc.}
\tablenotetext{c}{Position angle of the major axis, measured as positive 
moving eastward from north.}
\end{deluxetable}  

\clearpage
\begin{figure}
\epsscale{1.0}
\plotone{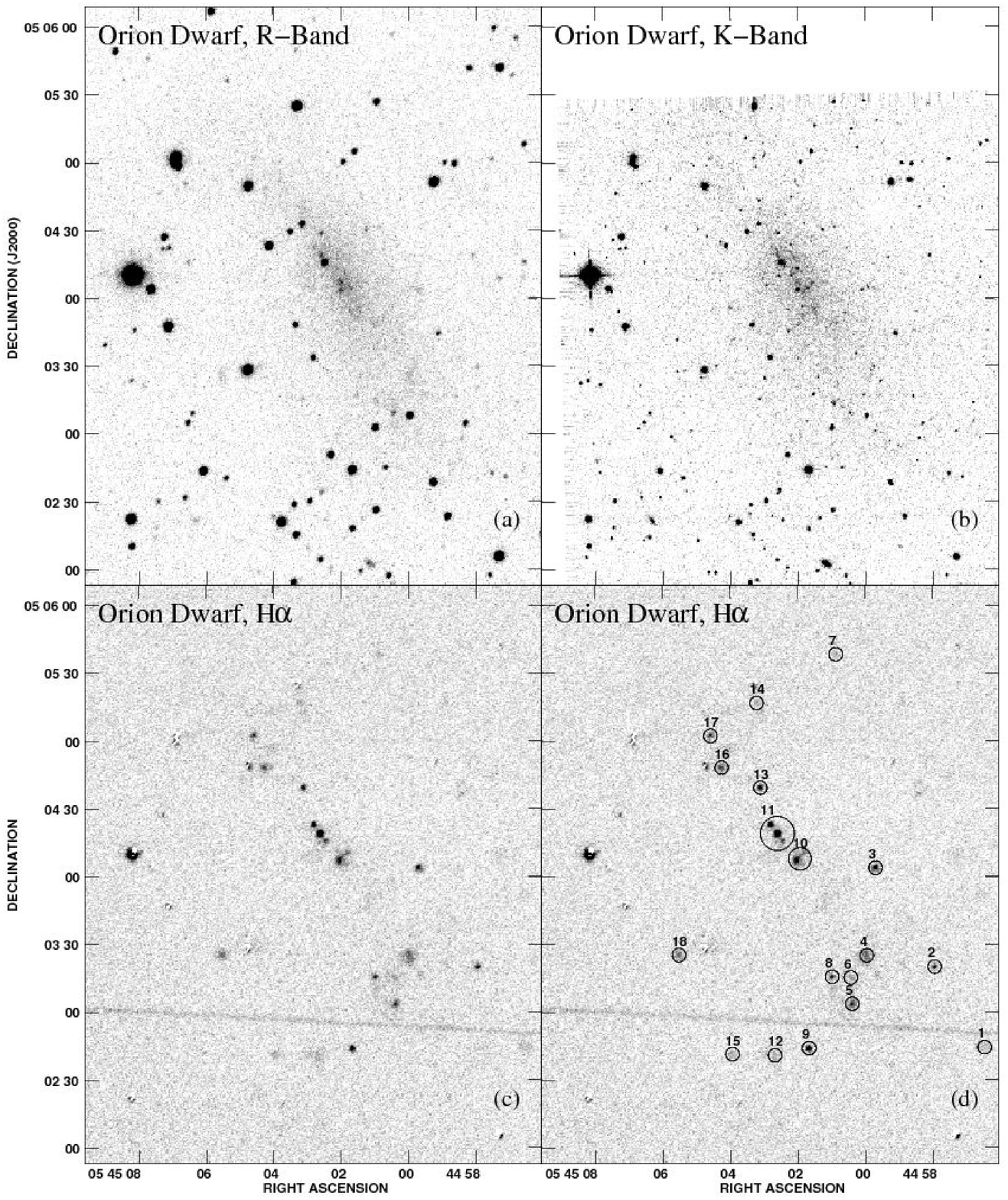}
\epsscale{1.0}
\caption{Optical and infrared images of the Orion dwarf.  Panel {\it
a} shows the R-band image from the {\it KPNO 0.9m} telescope; panel
{\it b} shows the \ks\ band image (courtesy of O. Vaduvescu); panel
{\it c} shows the continuum-subtracted \halpha\ image from the {\it
KPNO 0.9m} telescope; panel {\it d} shows the same image as in panel
{\it c}, but with the 18 individual \HII\ regions numbered in order of
increasing Right Ascension (note that Region\,\#11 consists of three
distinct peaks).}
\label{figcap1}
\end{figure}

\clearpage
\begin{figure}
\epsscale{0.8}
\plotone{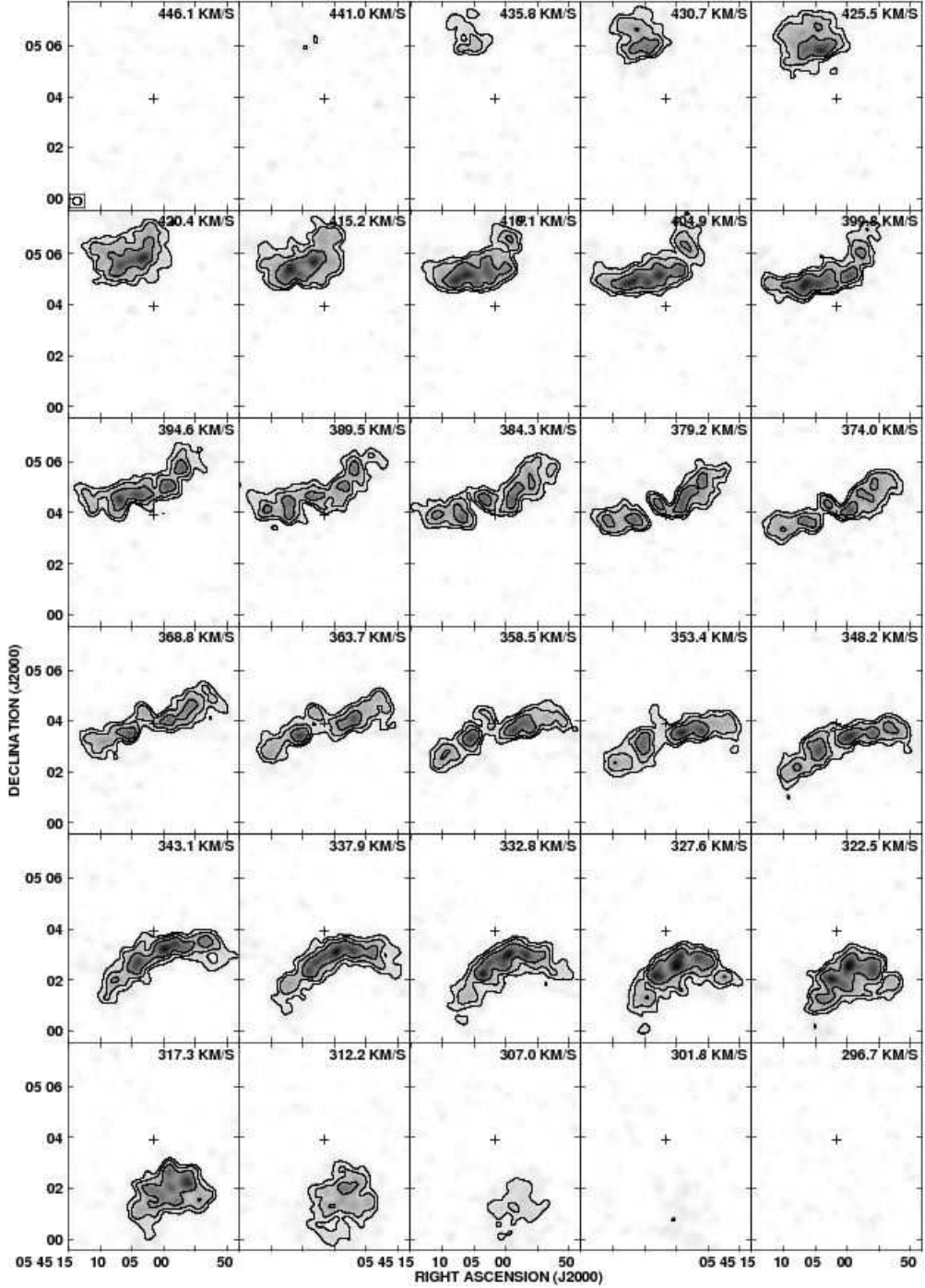}
\epsscale{1.0}
\caption{Channel maps from the 20\arcsec\ resolution datacube;
heliocentric velocity is noted in the upper right of each frame, and
the beam size is shown in the top left panel.  Contours are at levels
of 2.5, 5, 10 mJy\,Bm$^{-1}$.  Note the pronounced kinematic break in
the central regions (consider the 379.2 \kms\ panel and its
neighbors).  The $+$ sign marks the dynamical center as derived from
our tilted ring model fits (see \S~\ref{S3.1}).}
\label{figcap2}
\end{figure}

\clearpage
\begin{figure}
\epsscale{1.0}
\plotone{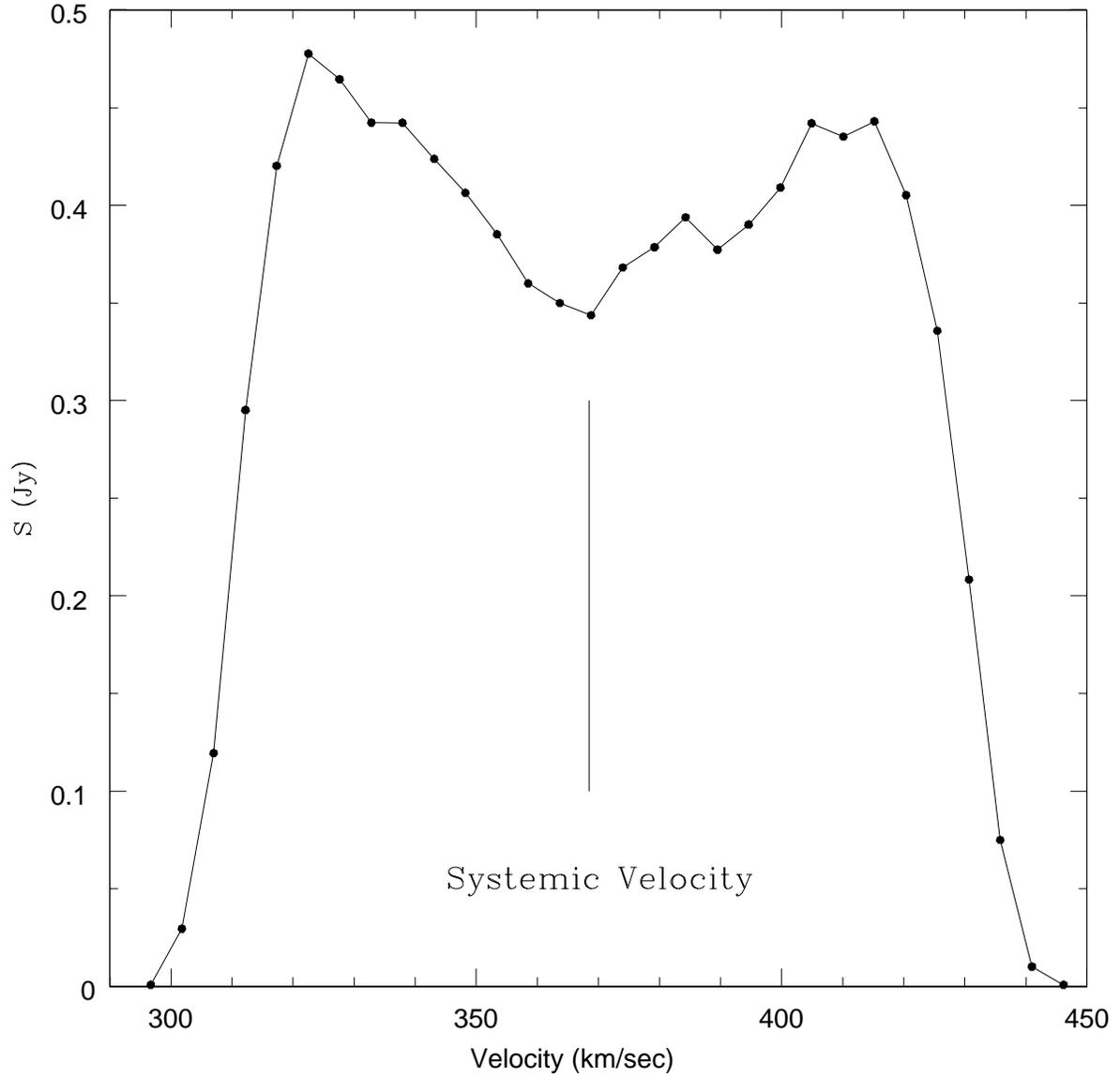}
\epsscale{1.0}
\caption{The global \HI\ profile of the Orion dwarf, created by
summing the flux in each channel of the low-resolution (20\arcsec\
beam), blanked data cube.  The systemic velocity, derived from this
profile and from our tilted ring analyses (see \S~{3.1}), is
368.5\,$\pm$\,1.0 \kms.}
\label{figcap3}
\end{figure}

\clearpage
\begin{figure}
\epsscale{0.93}
\plotone{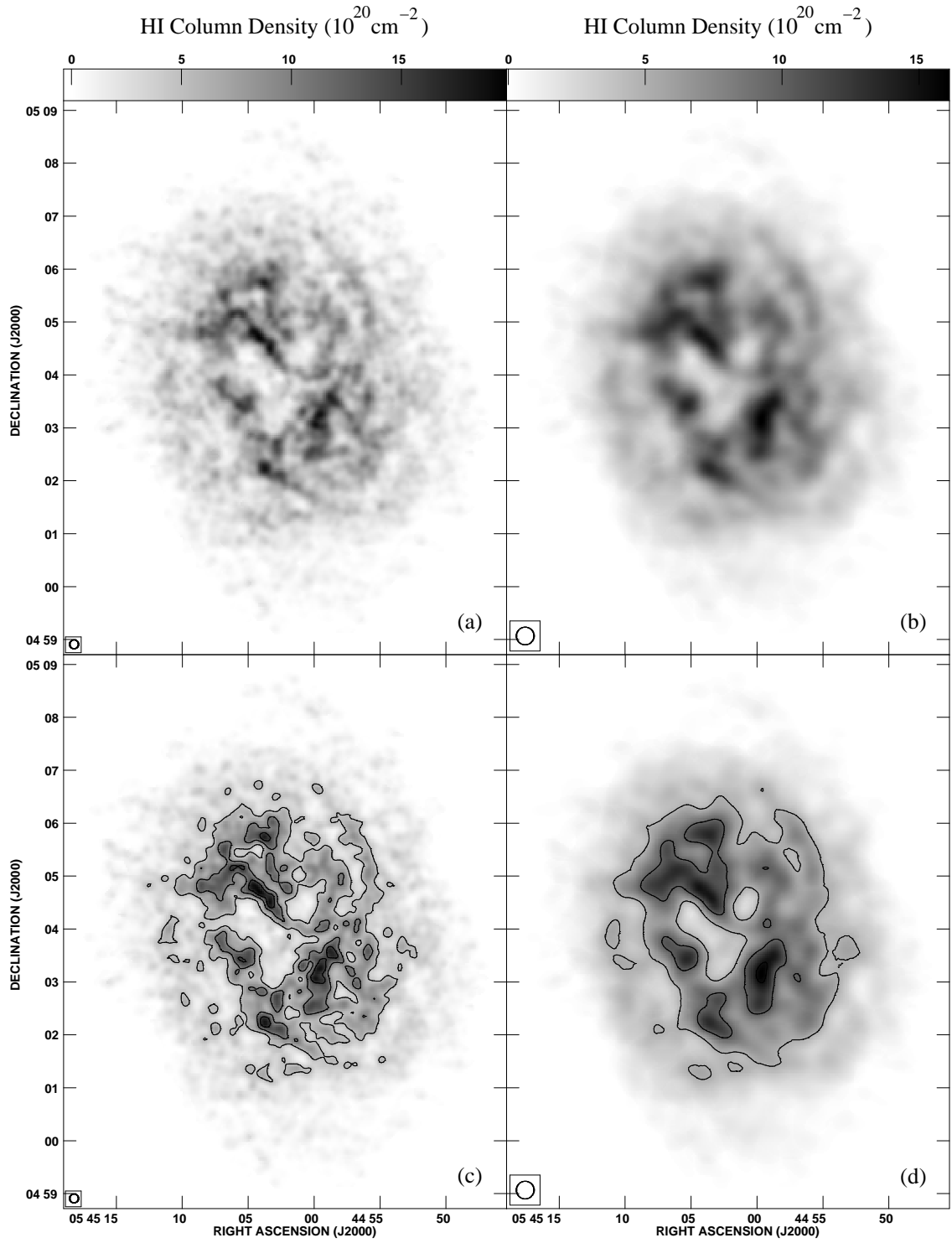}
\epsscale{1.0}
\caption{The \HI\ column density distribution of the Orion dwarf
galaxy at 10\arcsec\ (panels {\it a}, {\it c}) and 20\arcsec\ (panels
{\it b}, {\it d}) resolution.  Contours in the lower panels are shown
at the (5, 10, 15)\,$\times$\,10$^{20}$ cm$^{-2}$ levels.  A wealth of 
structure is prominent throughout the disk.}
\label{figcap4}
\end{figure}

\clearpage
\begin{figure}
\epsscale{0.8}
\plotone{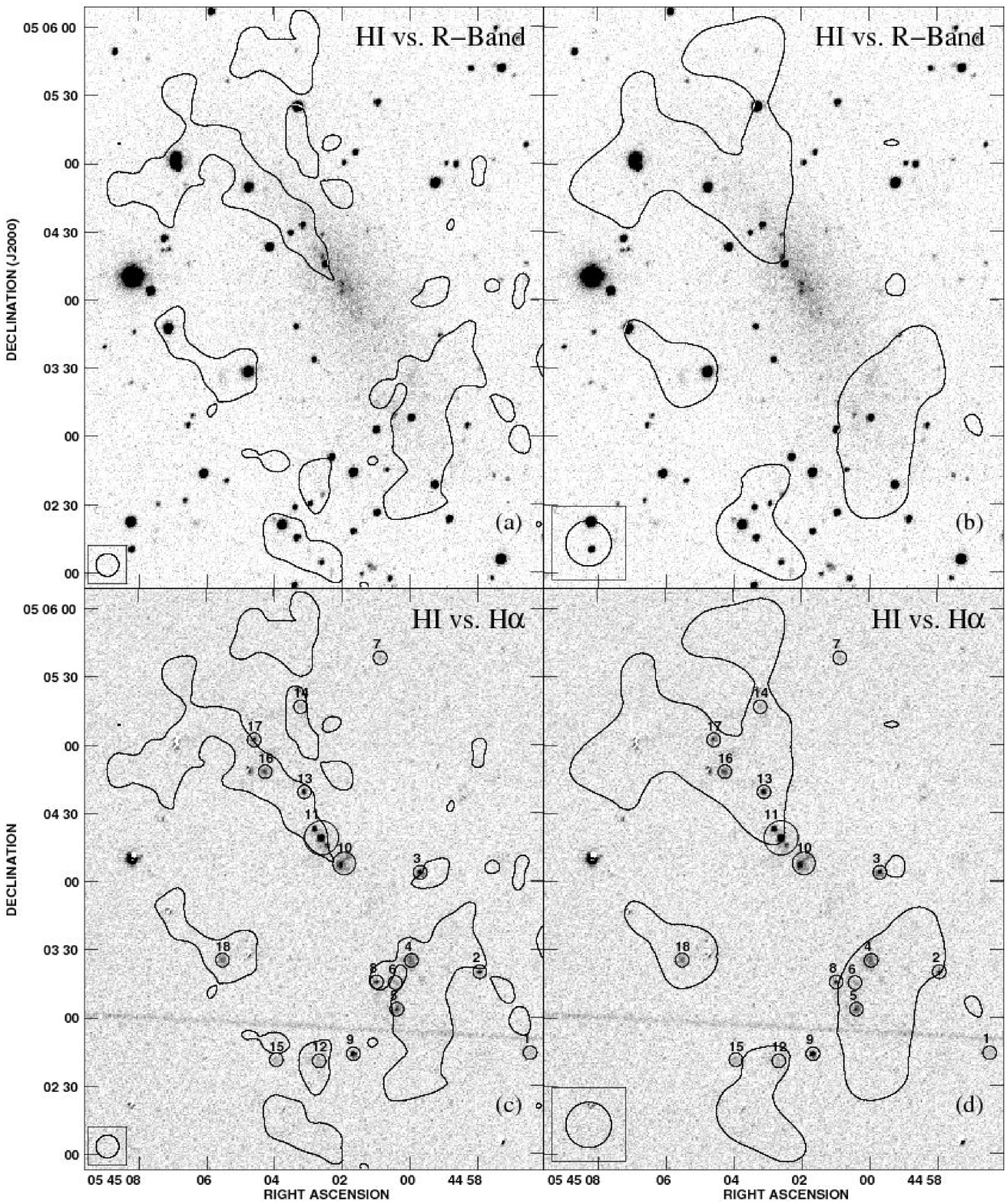}
\epsscale{0.8}
\caption{Optical images of the Orion dwarf, with the 10$^{-21}$
cm$^{-2}$ \HI\ column density contour superposed (beam size $=$
10\arcsec\ in panels {\it a} and {\it c}; beam size $=$ 20\arcsec\ in
panels {\it b} and {\it d}).  Panels {\it (a)} and {\it (b)} show the
stellar distribution; panels {\it (c)} and {\it (d)} show regions of
recent SF as traced by continuum subtracted \halpha\ emission.  We
detect 20 individual \HII\ regions (circled and numbered in panels
{\it c} and {\it d} in order of increasing Right Ascension).  Note
that \HII\ region \#11 consists of three distinct peaks that are just
barely resolved; the flux in Table 2 is the sum of these three.  The
regions of high \HI\ column density in general trace the locations of
active SF; this is in agreement with the ``canonical'' column density
threshold of 10$^{21}$ cm$^{-2}$ for active SF (e.g., {Skillman 1987},
{Kennicutt 1989}, {Kennicutt 1998b}, and references therein).}
\label{figcap5}
\end{figure}

\clearpage
\begin{figure}
\epsscale{1.0}
\plotone{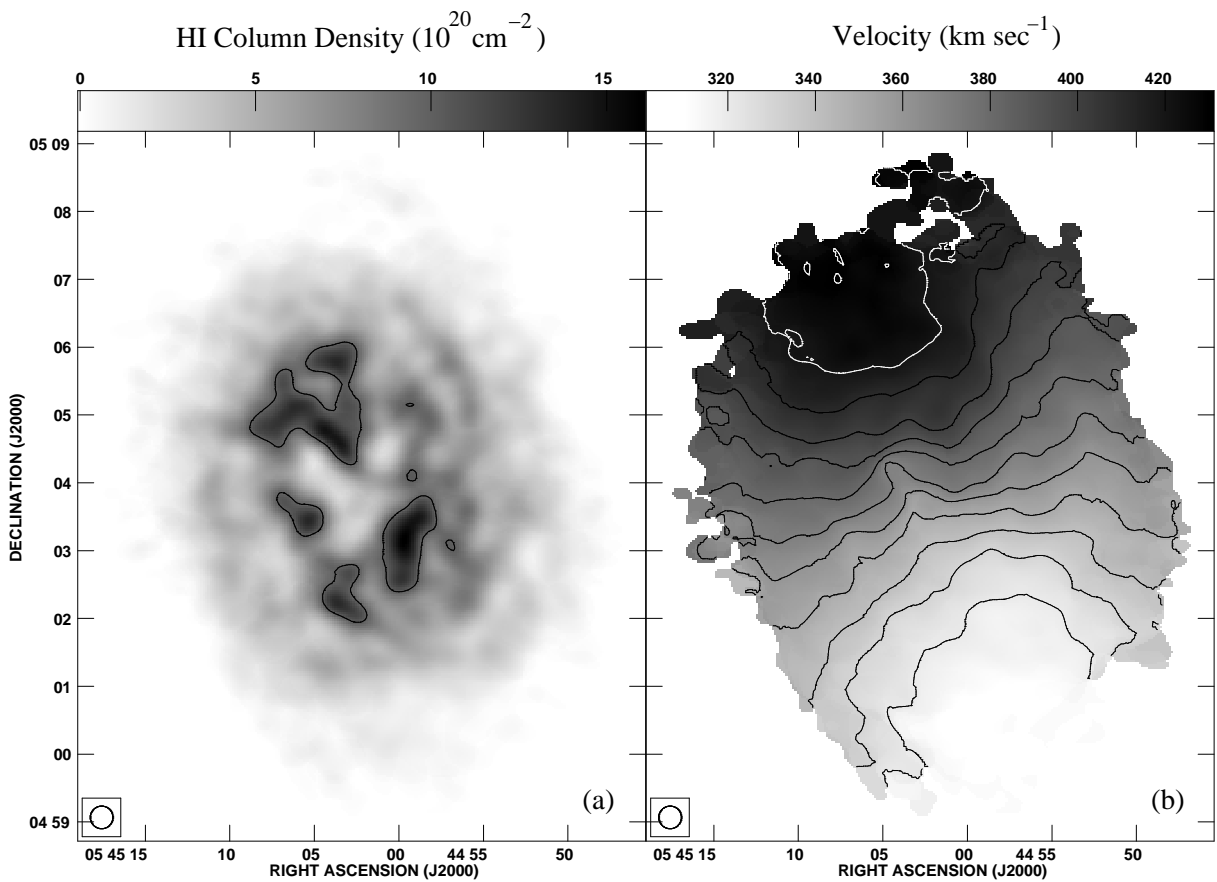}
\epsscale{1.0}
\caption{Comparison of the \HI\ column density distribution ({\it a})
and the integrated velocity field ({\it b}); the beam size is
20\arcsec.  The contour in {\it a} is at the 10$^{21}$ cm$^{-2}$ level
(c.f., Figures~\ref{figcap4}, \ref{figcap5}).  The contours in ({\it
b}) show velocities between 320 and 430 \kms, separated by 10 \kms.  The
Orion dwarf is undergoing well-ordered rotation throughout most of the
disk.  The inner regions have prominent ``hole/depression'' structures; a
pronounced kink in the velocity field is apparent at the same
location.}
\label{figcap6}
\end{figure}

\clearpage
\begin{figure}
\epsscale{1.0}
\plotone{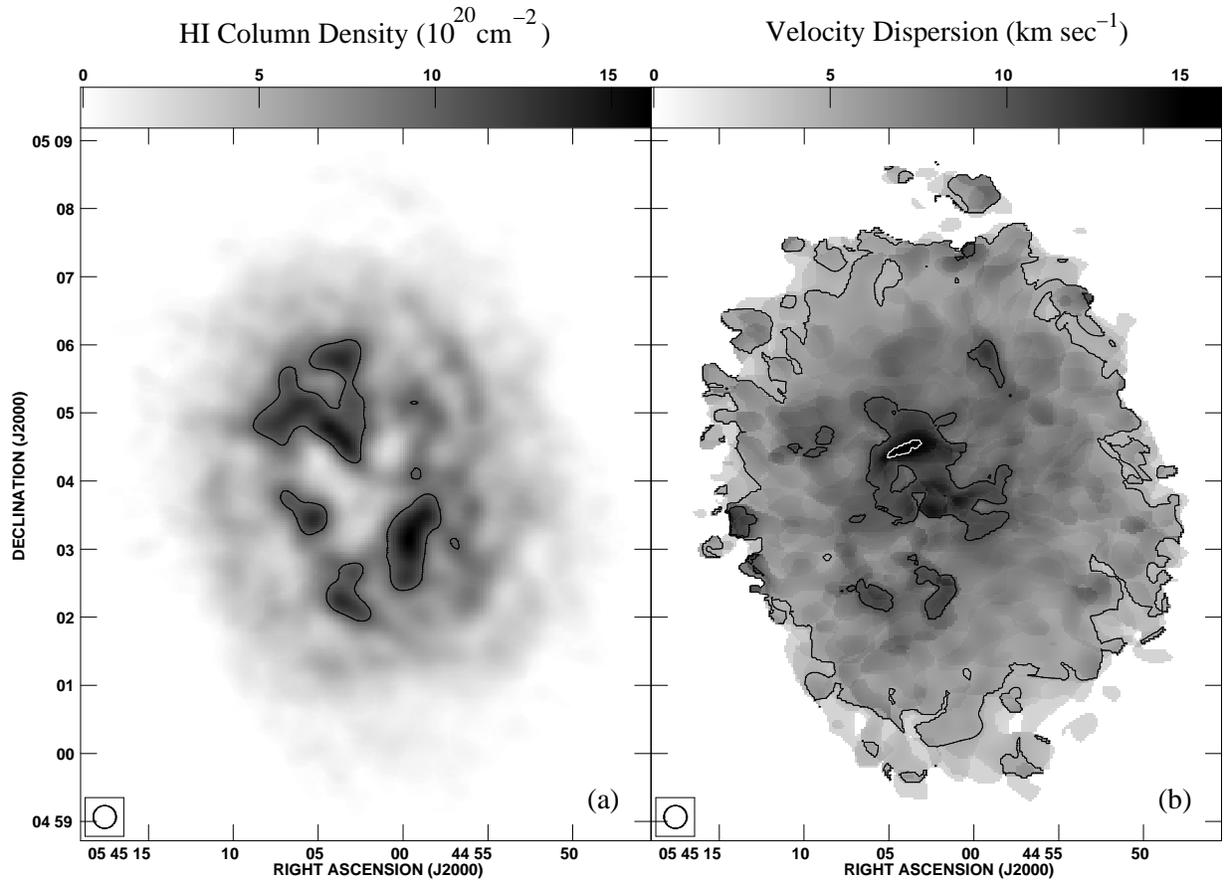}
\epsscale{1.0}
\caption{Comparison of the \HI\ column density distribution ({\it a})
and the intensity-weighted velocity dispersion ({\it b}); the beam
size is 20\arcsec.  The contour in ({\it a}) is at the 10$^{21}$
cm$^{-2}$ level (c.f., Figures~\ref{figcap4}, \ref{figcap5},
\ref{figcap6}); the contours in ({\it b}) are at the 5, 10, 15 \kms\
levels.  Note the prominent increase in velocity dispersion in the
inner disk.}
\label{figcap7}
\end{figure}

\clearpage
\begin{figure}
\epsscale{1.0}
\plotone{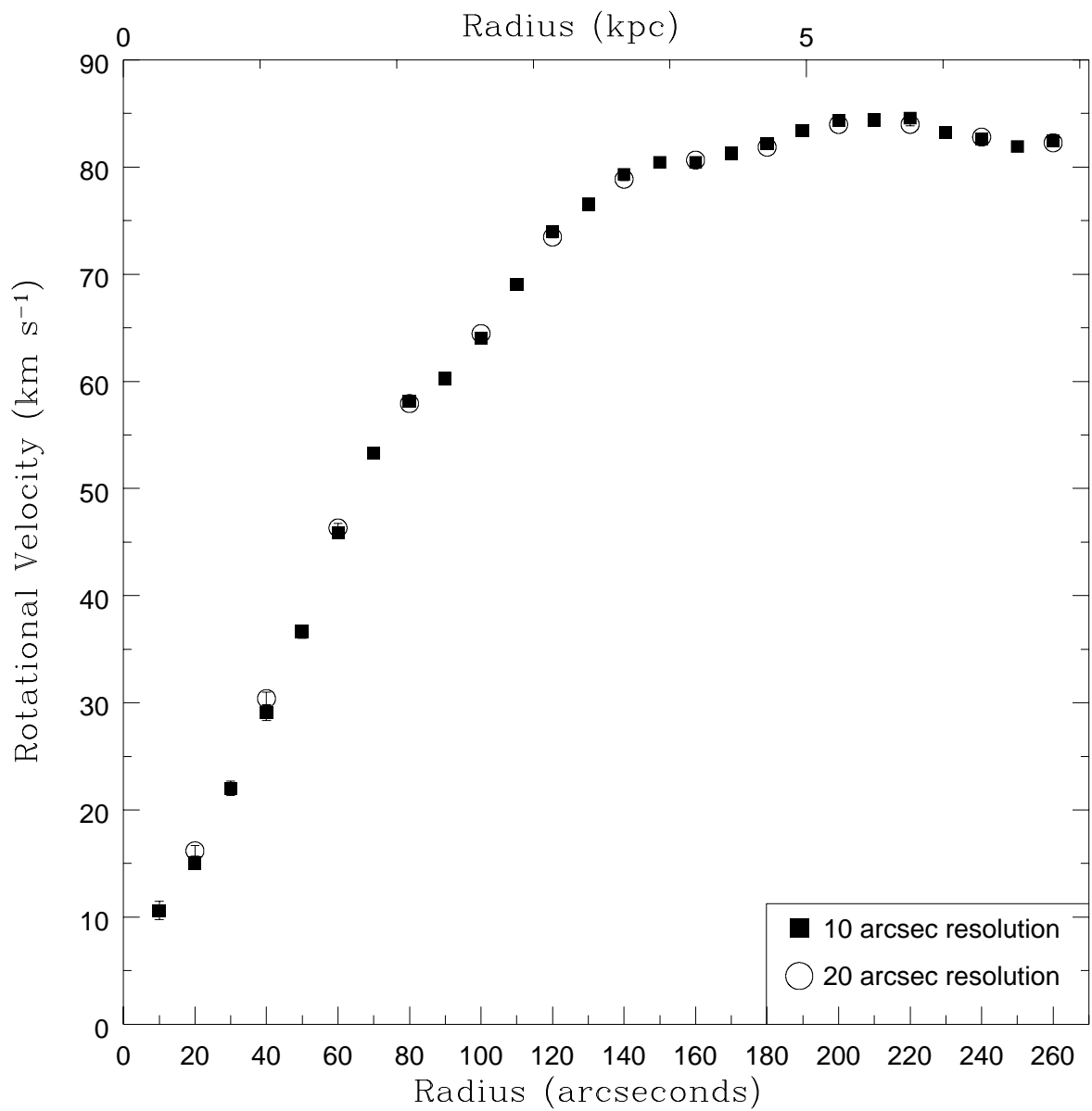}
\epsscale{1.0}
\caption{Rotation curve of the Orion dwarf galaxy, derived from both
the 10\arcsec\ and 20\arcsec\ datacubes; errorbars are plotted and in
most cases are smaller than the symbol sizes.  At a radius of
$\sim$140\arcsec\ ($\sim$3.7 kpc) the curve flattens to $\sim$82 \kms\
and remains flat to our detection limit.  The inferred dynamical mass
(M$_{\rm dyn}$ = 10.6$\times$10$^{9}$ \msun) exceeds the mass of the
gaseous and stellar components.}
\label{figcap8}
\end{figure}

\clearpage
\begin{figure}
\epsscale{1.0}
\plotone{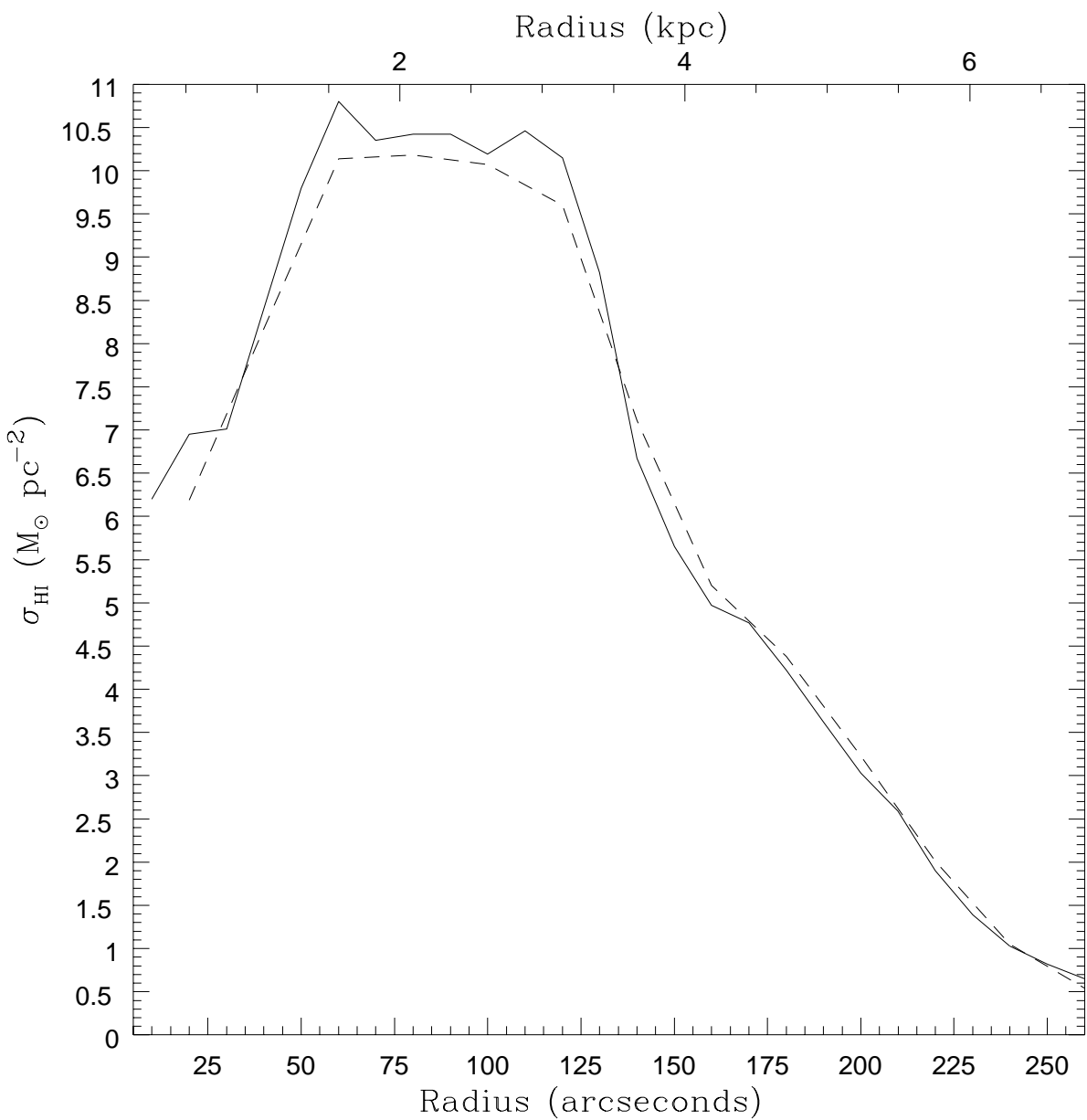}
\epsscale{1.0}
\caption{Radially averaged \HI\ mass surface density profiles of the
Orion dwarf, created by summing \HI\ emission in concentric rings
emanating from the dynamical center found in our rotation curve
analysis.  The solid line was created from the 10\arcsec\ resolution
image, while the dotted line was created from the 20\arcsec\ resolution
image.  These profiles shows that the inner ``holes/depressions''
have significantly less \HI\ gas mass per unit area than regions
at intermediate radii further out in the disk.}
\label{figcap9}
\end{figure}

\clearpage
\begin{figure}
\epsscale{1.0}
\plotone{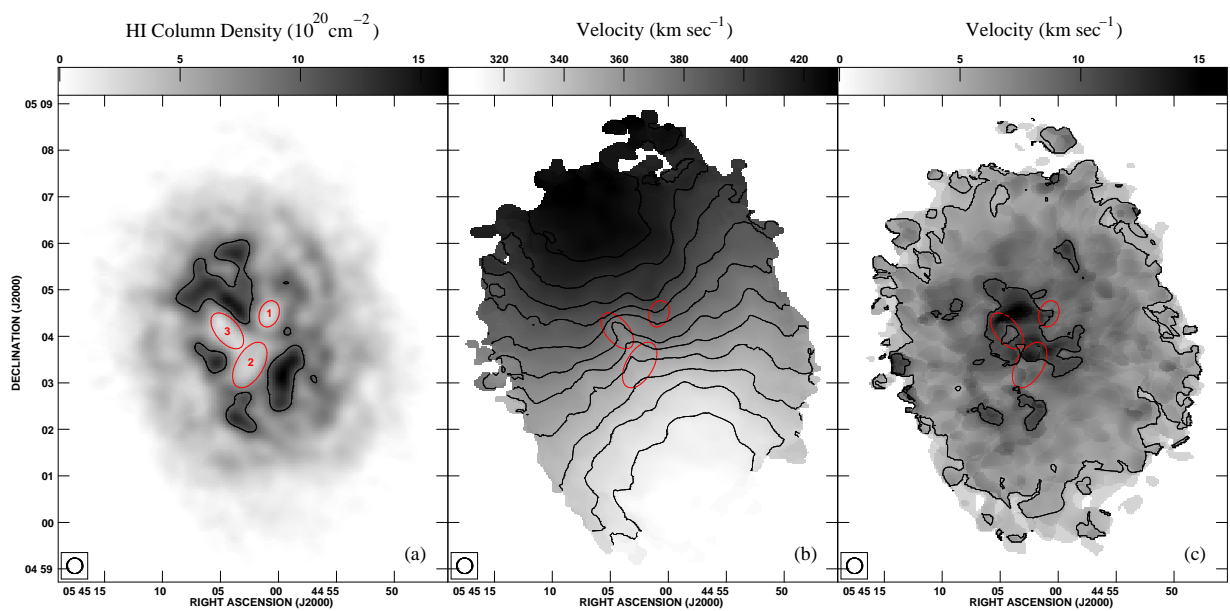}
\epsscale{1.0}
\caption{The \HI\ column density ({\it a}), velocity field ({\it b}),
and velocity dispersion ({\it c}) of the Orion dwarf at 20\arcsec\
resolution.  The contours in each panel are at the same levels as
those shown in Figures~\ref{figcap6} and \ref{figcap7}.  The red
ellipses shown denote the locations of the three most prominent
holes/depressions in the ISM.  Note that these features are
concentrated in the region of elevated velocity dispersion compared to
the more quiescent outer disk.  Further, the departure from ordered
rotation in the inner disk, prominent in ({\it b}), corresponds well
with the location of the easternmost of these features.}
\label{figcap10}
\end{figure}

\clearpage
\begin{figure}
\epsscale{1.0}
\plotone{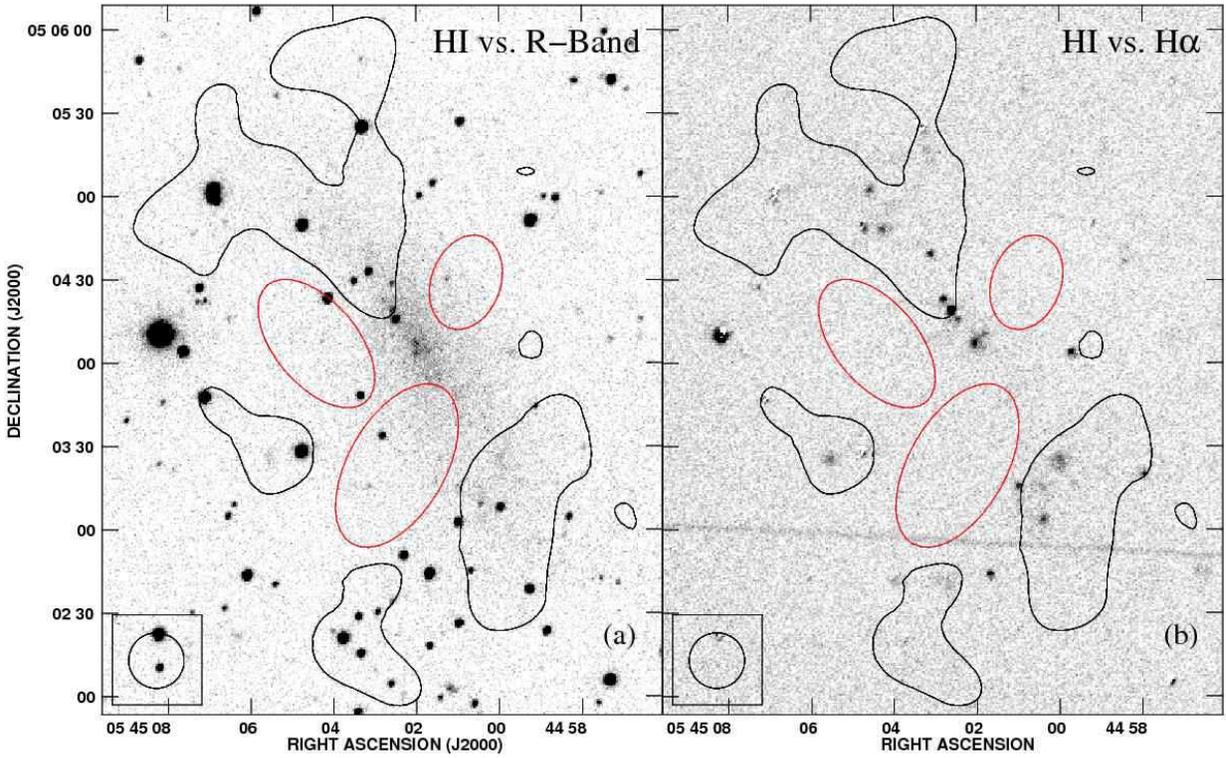}
\epsscale{1.0}
\caption{Locations of the three most prominent \HI\
``holes/depressions'' in the inner disk (shown as red ellipses;
c.f. Figures~\ref{figcap5} and \ref{figcap10}) compared to the stellar
({\it a}) and continuum subtracted \halpha\ ({\it b}) distributions.
The 10$^{21}$ cm$^{-2}$ \HI\ column density contour (20\arcsec\
resolution) is overlaid for reference in black.  Note that the central
stellar component occupies the region interior to these three
features.}
\label{figcap11}
\end{figure}

\clearpage
\begin{figure}
\epsscale{.80}
\plotone{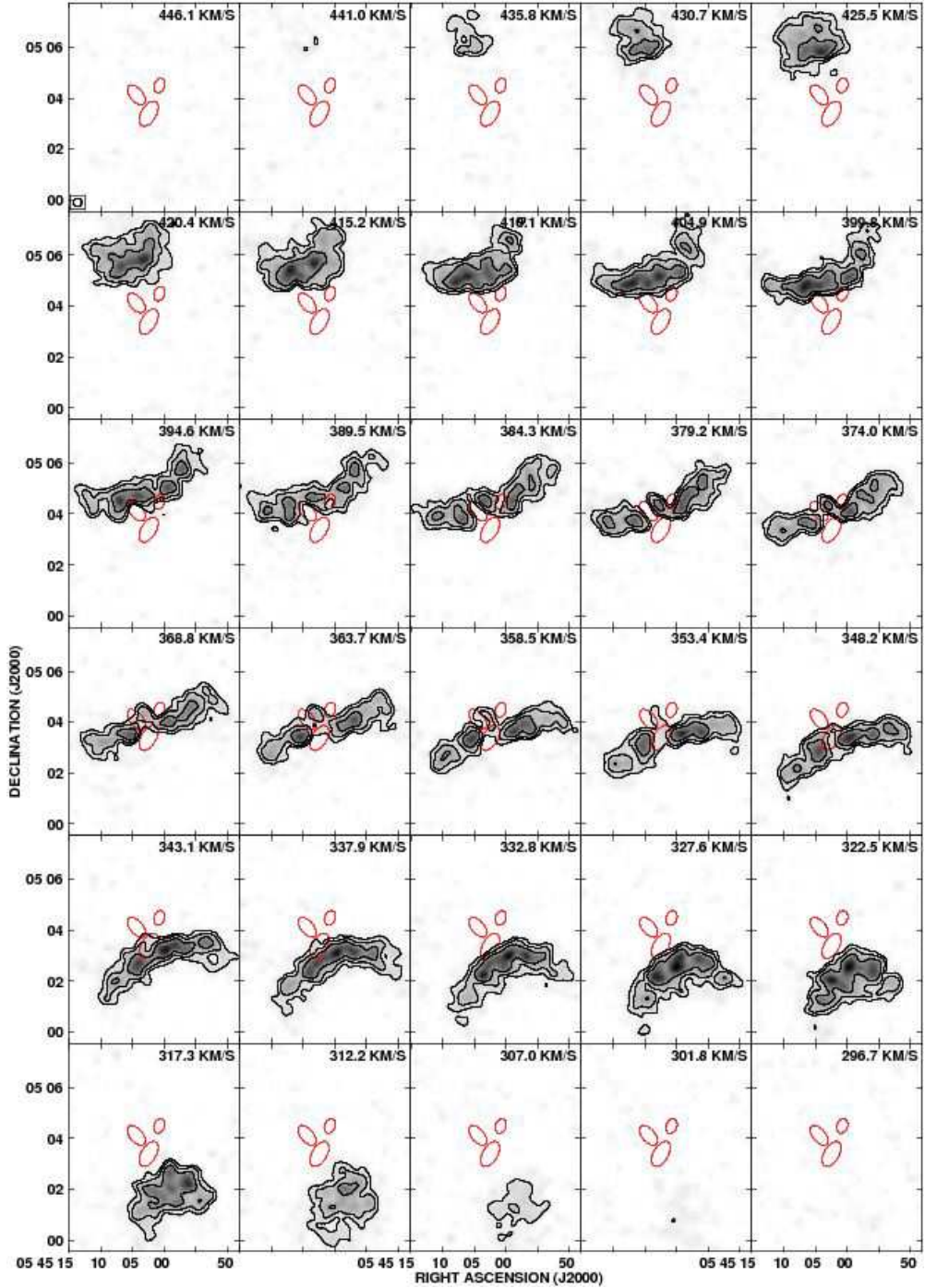}
\epsscale{1.0}
\caption{Same as Figure~\ref{figcap2}, except here the locations of
the three most prominent \HI\ ``holes/depressions'' are overlaid as
red ellipses (c.f., Figures~\ref{figcap10} and \ref{figcap11}).  Note
that while we do not detect the signatures of expansion in these
features at the current time, they are cleanly delineated in velocity
space (e.g., consider the easternmost ``hole/depression'' feature and
the void apparent in the panels centered around $\sim$380 \kms.)}
\label{figcap12}
\end{figure}

\clearpage
\begin{figure}
\epsscale{1.0}
\plotone{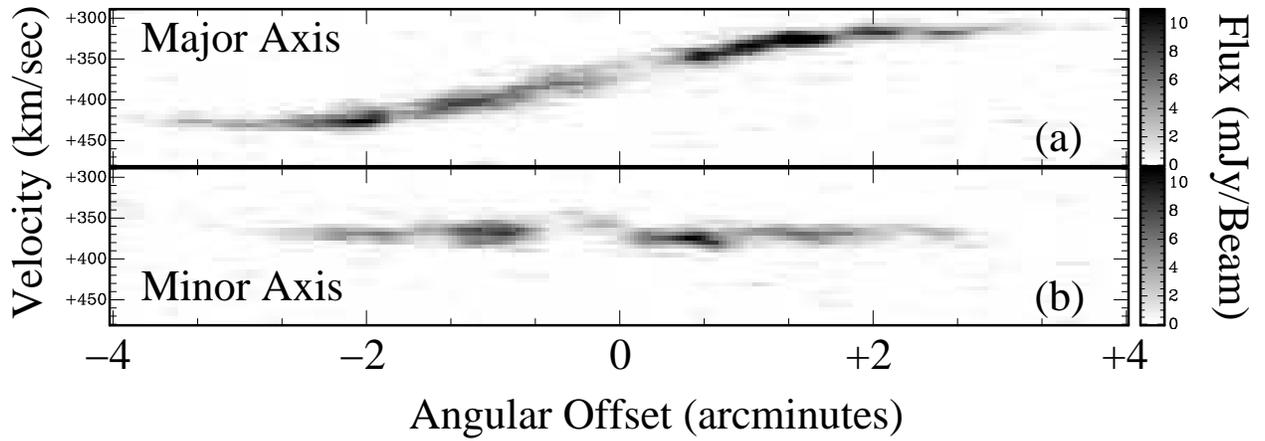}
\epsscale{1.0} 
\caption{Major (20\degree, measured east of north; {\it a}) and minor
(110\degree, measured east of north; {\it b}) axis position-velocity
diagrams.  Note the kinematic discontinuities arising from the
holes/depressions in the ISM [between 0 and $+$0.5\arcmin\ in ({\it a})
and between 0 and $-$0.5\arcmin\ in ({\it b})]; these can be compared
with Figures~\ref{figcap10} and \ref{figcap12}.  Each cut is centered
on the dynamical center (marked in Figure~\ref{figcap2}) of the galaxy
as derived from the rotation curve analysis (see
Figure~\ref{figcap8}).}
\label{figcap13}
\end{figure}

\end{document}